\documentclass[preprint,showpacs,showkeys,amsmath,amssymb,aps,doublespace]{revtex4}
\usepackage[dvips]{graphicx}
\usepackage{setspace}
\usepackage{amsmath}
\usepackage{amsfonts}
\usepackage{amssymb,color}
\usepackage{graphicx,color}
\usepackage{ifpdf}
\usepackage[latin1]{inputenc}
\usepackage{subfigure}

\begin{document}

\title{A traffic model with an absorbing-state phase transition}

\author{
M.~L.~L. Iannini,\footnote{email: lobao@div.cefetmg.br}
and
Ronald Dickman,\footnote{email: dickman@fisica.ufmg.br}
}

\address{
Departamento de F\'{\i}sica and
National Institute of Science and Technology for Complex Systems,\\
ICEx, Universidade Federal de Minas Gerais, \\
C. P. 702, 30123-970 Belo Horizonte, Minas Gerais - Brazil
}

\date{\today}

\begin{abstract}
We consider a modified Nagel-Schreckenberg (NS) model in which drivers do not decelerate if their
speed is smaller than the headway (number of empty sites to the car ahead).  (In the original NS model,
such a reduction in speed occurs with probability $p$, independent of the headway, as long as the current speed is greater than zero.)
In the modified model the free-flow state
(with all vehicles traveling at the maximum speed, $v_{max}$) is
{\it absorbing} for densities $\rho$ smaller than a critical value $\rho_c = 1/(v_{max} + 2)$.
The phase diagram in the $\rho - p$ plane is reentrant:
for densities in the range $\rho_{c,<} < \rho < \rho_c$, both small and large values of $p$ favor free flow,
while for intermediate values, a nonzero fraction of vehicles have speeds $< v_{max}$.
In addition to representing a more realistic description of driving behavior, this change leads to
a better understanding of the phase transition in the original model.  Our results
suggest an unexpected connection between traffic models and stochastic sandpiles.
\end{abstract}

\maketitle

\section{Introduction}

The Nagel-Schreckenberg (NS) model holds a central position in traffic modeling via cellular automata, because it
reproduces features commonly found in real traffic, such as the transition between free flow to jammed state,
start-and-stop waves, and shocks (due to driver overreaction).
This simple model represents the effect of fluctuations in driving behavior by
incorporating a stochastic element: the spontaneous reduction of velocity with probability $p$.

Although the NS model has been studied extensively, the nature of the transition
between free and jammed flow, in particular, whether it corresponds to a critical point, remains controversial \cite{re,ft1,ft2,ft3}.
Modifications in update rules of the NS model have been found to result in a well defined phase transition \cite{ft4}.
Krauss \emph{et al.} \cite{ft5} proposed a generalized version of the NS model and showed numerically that
free- and congested-flow phases may coexist.
While the NS model does not exhibit metastable states, which are important in
observed traffic flow, including a slow-to-start rule, such that
acceleration of stopped or slow vehicles is delayed compared to that of moving or faster cars,
can lead to metastability \cite{hist1,hist2,hist3}.
Takayasu and Takayasu \cite{hist1} were the first to suggest a cellular automaton (CA)
model with a slow-to-start rule. Benjamin, Johnson, and Hui introduced a different slow-to-start rule in Ref. \cite{hist2},
while Barlovic \emph{et al.} suggested a velocity-dependent randomization model \cite{hist3}.
Other models with metastable
states are discussed in Refs. \cite{hist4,hist5}. A review of CA traffic models is presented in Ref. \cite{book}.

In the original NS model, at each time step (specifically, in the reduction substep), a driver with nonzero velocity
reduces her speed with probability $p$.
Here we propose a simple yet crucial modification, eliminating changes in speed in this substep
when the distance to the car ahead is greater than the current speed.  We believe that this rule
reflects driver behavior more faithfully than does the original reduction step, in which drivers may decelerate for no
apparent reason.  While one might argue that distractions such as cell phones cause drivers to decelerate
unnecessarily, we can expect that highways will be increasingly
populated by driverless vehicles exhibiting more rational behavior.
The modified model, which we call the absorbing Nagel-Schreckenberg (ANS) model, exhibits a line of
absorbing-state phase transitions between free and congested flow in the $\rho-p$ plane. (Here $\rho$ denotes
the {\it density}, i.e., the number of vehicles per site.)
The modification proposed here allows us to understand the nature of the phase transition in the original model,
and to identify a proper order parameter.  The ANS model exhibits a surprising reentrant phase diagram.

Regarding the nature of the phase transition in the original NS model,
the key insight is that, for $p=0$, it exhibits a transition between an absorbing state (free flow)
and an active state (congested flow) at density $\rho = 1/(v_{max}+1)$, where $v_{max}$ denotes the
maximum speed.  Free flow is absorbing because each car advances the
same distance in each time step, so that the configuration simply executes rigid-body motion (in the co-moving frame it
is frozen).  Congested flow, by contrast, is active in the sense that the distances between vehicles
change with time.
Below the critical density, activity (if present initially) dies out, and an absorbing
configuration is reached; above the critical density there must be activity, due to lack of sufficient space between vehicles.
Setting $p>0$ in the original model is equivalent to including a {\it source} of spontaneous activity. Since such a source
eliminates the absorbing state \cite{marro}, the original NS model {\it does not possess a phase transition for $p>0$}.
(It should nonetheless be possible to observe scaling phenomena as $p \to 0$.) A similar conclusion was reached by
Souza and Vilar \cite{ft3}, who drew an analogy between the phase transition at $p=0$ and a quantum phase
transition at temperature $T=0$.
In their analogy, $p>0$ corresponds to $T>0$, for which, {\it sensu stricto}, there is again no phase transition.

The remainder of this paper is organized as follows. In the next section we define the ANS model, pointing out how it differs from the
original NS model.  In Sec. III we explain qualitatively the nature of the phase diagram, and report simulation
results for the phase boundary.  Sec. IV presents results on critical behavior, followed in Sec. V by a summary and discussion
of our findings.

\section{model}

The NS model and its absorbing counterpart (ANS) are defined on a ring of $L$ sites, each of which may be empty or occupied by a vehicle with
velocity $v=0,1,...,v_{max}$.  (Unless otherwise noted, we use $v_{max}=5$, as is standard in studies of the NS model.)
The dynamics, which occurs in discrete time, conserves the number $N$ of vehicles; the associated
intensive control parameter is $\rho = N/L$.  Denoting the position of the $i$-th vehicle by $x_i$, we define the headway
$d_i =x_{i+1}-x_{i}-1$ as the number
of empty sites between vehicles $i$ and $i+1$.  Each time step consists of four substeps, as follows:

 \begin{itemize}
 \item Each vehicle with $v_{i}<v_{max}$ increases its velocity by one unit: $v_{i}\rightarrow v_{i}+1$
 \item Each vehicle with $v_{i}>d_{i}$ reduces its velocity to $v_{i}=d_{i}$.
 \item NS model: each vehicle reduces its velocity by one unit with probability $p$.\\
 $\;$ ANS model: each vehicle with $v_i \!=\! d_i$ reduces its velocity by one unit with probability $p$.
 \item All vehicles advance their position in accord with their velocity.
 \end{itemize}

In practice, given the velocities $v_i$ and headways $d_i$, there is no need to keep track of positions:
the final substep is simply $d_i \to d_i - v_i + v_{i+1}$ for $i=1,...,N-1$, and $d_N \to d_N - v_N + v_1$.

\begin{figure}[!htb]
\includegraphics[clip,angle=0,width=0.4\hsize]{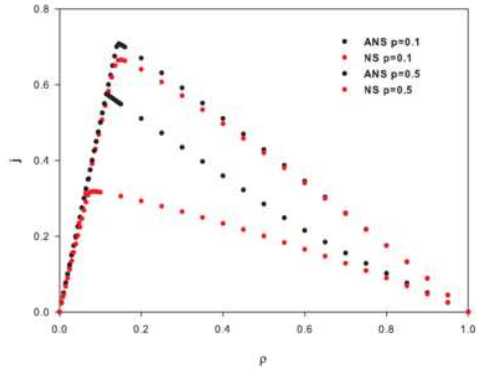}
  \caption{\label{comp} (Color online) Flux $j$ versus density in the NS and ANS models for probabilities $p=0.1$ (upper)
  and $p=0.5$ (lower). System size $L=10^5$; vehicles are distributed randomly at $t=0$. Error bars are smaller than symbols.}
\end{figure}

The modification of the third substep leads to several notable changes in behavior, as reflected in the fundamental diagram shown in Fig.~\ref{comp}.  Evidently, there is a phase transition in the ANS model for $p>0$, while there is none in the NS model.
The flux $q$ generally takes its maximum value at the transition. (For small $p$, however,
maximum flux occurs at a density above $\rho_c = 1/(v_{max}+2)$,  approaching $\rho=\frac{1}{v_{max}+1}$ for $p=0$).
The low-density, absorbing phase has $v_i = v_{max}$ and $d_{i} \geq v_{max}+1, \; \forall i$;
in this phase all drivers advance in a deterministic manner, with the flux given by $J=\rho v_{max}$.
In the active state, by contrast,
a nonzero fraction of vehicles have $d_i \leq v_{max}$.  For such vehicles, changes in velocity are possible, and the configuration
is nonabsorbing.
The stationary fluxes in the NS and ANS models differ significantly
over a considerable interval of densities, especially for high values of $p$. Below the critical density $\rho_{c}$, this difference is due the existence of an absorbing phase in the ANS model. For densities slightly above $\rho_{c}$, most vehicles
have velocity $v_i = v_{max}$ and $d_i = v_{max}+1$, although there is no absorbing state.
As the density approaches unity, the differences between the fluxes in the ANS and NS models become smaller.

\begin{figure}[h]
\center
\subfigure[]{\includegraphics[clip,angle=0,width=0.4\hsize]{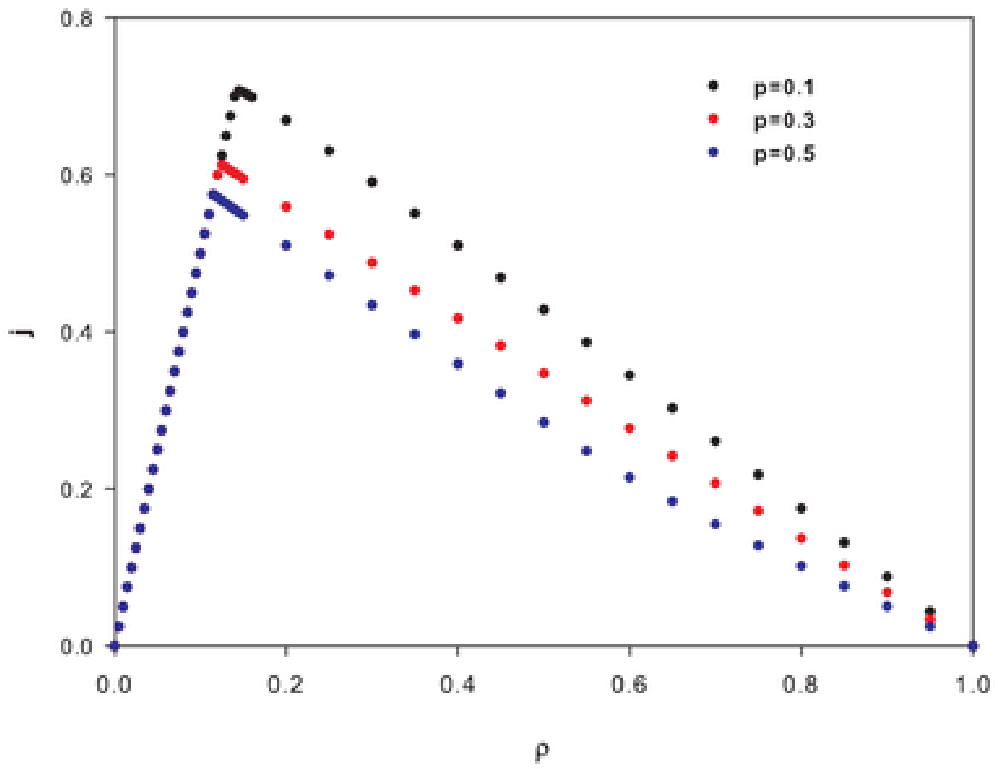}}
\qquad
\subfigure[]{\includegraphics[clip,angle=0,width=0.4\hsize]{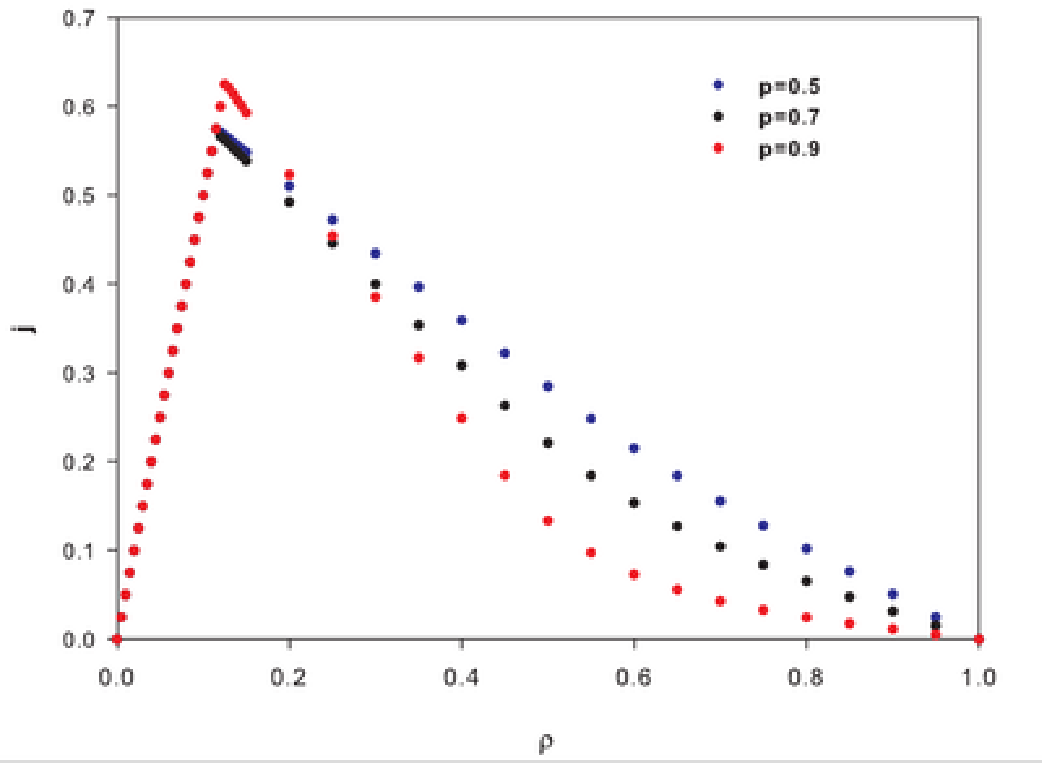}}
\caption{ \label{13579} Flux versus density in the ANS model for (a) $p=0.1$, $0.3$ and $0.5$, and (b) $p=0.5$, $0.7$ and $0.9$.
Note that the density of maximum flux first decreases, and then increases, with increasing $p$; the minimum occurs near $p\simeq 0.5$.
System size $L=10^5$; error bars are smaller than symbols.}
\end{figure}

For fixed deceleration probability $p$, the flux $j = \rho \overline{v}$ first grows, and then decreases as we increase the vehicle density $\rho$.
An intriguing feature is the dependence of the density at maximum flux on the probability $p$: Fig.~\ref{13579} shows that the
density at maximum flux decreases with increasing $p$ until reaching a minimum near $p=0.5$, and subsequently increases with increasing $p$.
This reflects the reentrant nature of the phase diagram, as discussed in Sec.~\ref{phasediagram}.

\subsection{Special cases: $p=0$ and $p=1$}

For the extreme values $p=0$ and $p=1$ the ANS model is deterministic; these two cases deserve comment.
For $p=0$, the NS and ANS models are identical.  The system reaches an absorbing state, $v_i=v_{max}$, $\forall i$,
for densities $\rho \leq 1/(v_{max}+1)$.  For higher densities we observe nonzero activity in the steady state.
We note however that there are special configurations, in which $v_i = d_i$, $\forall i$, with some $v_i < v_{max}$,
whose evolution corresponds to a rigid rotation of the pattern.  (A simple example is $v_i = d_i = n$, $\forall i$, with $n=1$, 2, 3 or 4,
and density $\rho = 1/(n+1)$.)  Since our interest here is in the model with $0 < p < 1$ we do not comment further
on such configurations.

For the NS model with $p=1$, from one step to the next, each velocity $v_i$ is nonincreasing.  (Of course $v_i \to v_i + 1$ at
the acceleration substep, but this is immediately undone in the subsequent substeps.)
Thus if the evolution leads to a state in which even one vehicle has velocity zero, all vehicles eventually stop.
Such an event is inevitable for $\rho > 1/3$, since in this case $d_i \leq 1$ for at least one vehicle, which is obliged to
have $v_i = 0$ after one step.
For $\rho\leq\frac{1}{3}$, steady states with nonzero flux are possible, depending on the choice of initial
condition.
Such configurations are {\it metastable}
in the sense that the stationary state depends on the initial distribution.
In the ANS model with $p=1$ the mean velocity in steady state is zero only for $\rho \geq 1/2$.
For  $\rho \leq 1/(v_{max}+2)$,
we find that the system always reaches an absorbing configuration with $\overline{v}= v_{max}$.  In the remaining
interval, $1/(v_{max}+2) < \rho \leq 1/2$, we find $\overline{v}= 1 - 2\rho$.

\section{Phase diagram}
\label{phasediagram}

\subsection{Initial condition dependence}

In studies of traffic, states are called {\it metastable} if they can be obtained from some, but not all initial
conditions \cite{hist1,hist2,hist3,hist4,hist5};
such states are an essential component of real traffic.
Since the NS model is not capable of reproducing this feature, models with modified update rules have been investigated by
several authors \cite{hist1,hist2,hist3}.
In the ANS model, by contrast, there is a region in the $\rho - p$ plane in which, depending on the initial condition,
the system may evolve to an active state or an absorbing one.  Our results are consistent with the usual scenario for
absorbing-state phase transitions \cite{marro,odor07,henkel}: activity in a finite system has a finite lifetime;
in the active phase, however, the mean lifetime diverges as the system size tends to infinity.  Properties of the
active phase may be inferred from simulations that probe the {\it quasistationary regime} of large but finite systems \cite{qssim}.

To verify the existence of metastable states in the ANS model, we study its evolution starting from two very different classes of initial conditions (ICs): homogeneous and jammed. In a homogeneous IC, the headways $d_i$ are initially are uniform as possible, given the density
$\rho = 1/(1+ \overline{d})$, where $\overline{d}$ denotes the mean headway. In this case the initial velocity is $v_{max}$ for all vehicles.
In a jammed IC, $N$ vehicles occupy the $N$ contiguous sites, while the remaining $N(\rho^{-1} - 1)$ sites
are vacant; in this case $d_i=0$ for $i = 1,...,N-1$, and only vehicle $N$ has a nonzero initial velocity ($v_N = v_{max}$).
Homogeneous ICs are much closer to an absorbing configuration than are jammed ICs.

\begin{figure}[h]
  \centering
  \includegraphics[clip,angle=0,width=0.4\hsize]{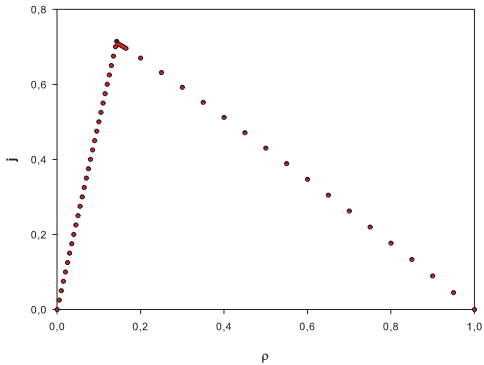}
  \caption{Steady-state flux versus density for $p=0.5$ and $L=10^{5}$.  Homogeneous and jammed ICs lead to identical stationary states except for a small interval of densities near maximum flux. Error bars are smaller than symbols.}
  \label{flux1}
\end{figure}

\begin{figure}[h!]
  \centering
  \includegraphics[clip,angle=0,width=0.4\hsize]{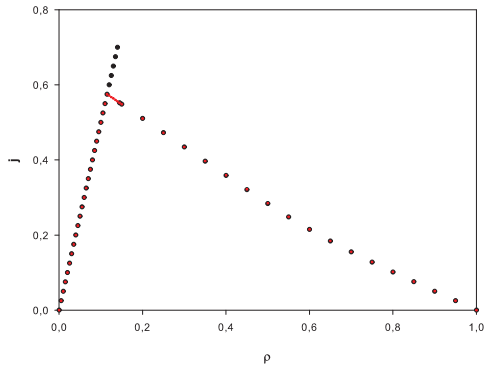}
  \caption{Steady-state flux versus density as in Fig.~\ref{flux1}, but for $p=0.5$.
  Homogeneous and jammed ICs curves lead to identical stationary states
   except for densities near maximum flux.}
  \label{flux5}
\end{figure}

Figure~\ref{flux1} shows the fundamental diagram obtained using homogeneous and jammed ICs for $p=0.1$; for this value of $p$ the stationary state is the same, regardless of the IC, except near $\rho=\frac{1}{7}$ where, for the homogeneous ICs, an absorbing configuration is attained,
having a greater steady-state flux than obtained using jammed ICs.
For higher probabilities $p$, we find a larger interval of densities in which the stationary behavior depends in the choice of IC.
In Fig.~\ref{flux5}, for $p=0.5$, this interval corresponds to  $0.118 \leq \rho \leq 0.143$; higher fluxes (black points) are obtained
using homogeneous ICs, and lower fluxes (red) using jammed ICs.  Homogeneous ICs rapidly evolve to an absorbing configuration, while
jammed ICs, which feature a large initial activity, do not fall into an absorbing configuration for the duration of the simulation
($t_{max}=10^7$), for the system size ($L=10^5$) used here.

Systematic investigation of the steady-state flux obtained using homogeneous and jammed ICs leads to the conclusion that the $\rho$ - $p$
plane can be divided into three regions.  To begin, we recall that for $\rho > \rho_c = 1/(v_{max} + 2)$ and $p>0$, the
mean velocity $\overline{v}$ must be smaller than $v_{max}$.  Thus the activity is nonzero and the configuration (i.e.,
the set of values $v_i$ and $d_i$) changes.  In this region, homogeneous and jammed ICs always lead to the same steady state.

\begin{figure}[h!]
  \centering
  \includegraphics[clip,angle=0,width=0.4\hsize]{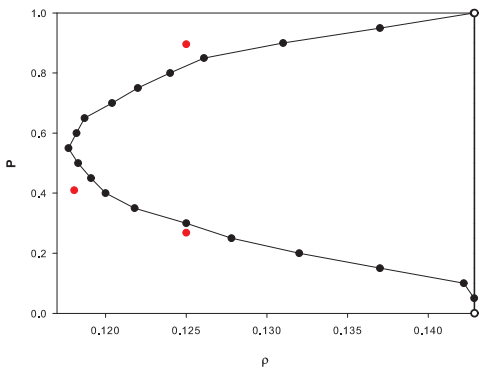}
  \caption{Boundary between active and absorbing phases in the $\rho$ - $p$ plane.  Black points: preliminary estimates
  from initial-condition dependence as explained in text.  Red points: precise estimates obtained via finite-size
  scaling as described in Sec.~IV.  The open circle at $\rho=1/7$, $p=0$ is not part of the
  phase boundary: for $p=0$ the transition occurs at $\rho = 1/6$.  The open circle $\rho=1/7$, $p=1$ marks the
  other end of the phase boundary; we note however that at this point, all initial conditions evolve to the
  absorbing state.}
  \label{tc}
\end{figure}

For $\rho < \rho_c$, absorbing configurations exist for any value of $p$.  There is nevertheless a region with $\rho < \rho_c$
in which activity is long-lived.  In this region, which we call the {\it active phase},
the steady state depends on whether the IC has little activity (homogeneous) or
much activity (jammed).  Outside this region, all ICs evolve to an absorbing configuration; we call this the
{\it absorbing phase}.  The boundary between the active and absorbing phases, determined via the criterion of
different steady states for homogeneous and jammed ICs,
is shown in Fig.~\ref{tc}.

Our results are consistent with the following scenario, familiar from the study of phase
transitions to an absorbing state \cite{marro,odor07,henkel}: for finite systems, all ICs with $\rho < \rho_c$ and $p>0$ eventually fall into an
absorbing configuration.  Within the active phase, however, the mean lifetime of activity grows exponentially with system size.
The phase boundary represents a line of critical points, on which the lifetime grows as a power law of system size.
(Further details on critical behavior are discussed in Sec.~\ref{critical}.)  A surprising feature of the phase boundary is that it is
{\it reentrant}: for a given density in the range $0.116 < \rho < \rho_c$, the absorbing phase is observed for both small and large $p$ values,
and the active phase for intermediate values.  The reason for this is discussed in Sec.~III.C.  We denote the upper and lower
branches of the phase boundary by $p_+(\rho)$ and $p_-(\rho)$, respectively; they meet at $\rho_{c,<} \simeq 0.116$.

The phase boundary is singular at its small-$p$ limit. As $p$ tends to zero from positive values, the critical
density approaches 1/7, but for $p=0$ the transition occurs at $\rho = 1/6$.  The phase diagram of the ANS model for $0 < p < 1$
is similar to that of a stochastic sandpile \cite{granada,pruessner}.  In the sandpile, there are no absorbing configurations for
particle density $\rho > z_c - 1$, where $z_c$ denotes the toppling threshold; nevertheless, the absorbing-state phase transition
at a density strictly smaller than this value.  Similarly, in the ANS model there are no absorbing configurations for $\rho > 1/7$,
but the phase transition occurs at some smaller density, depending on the deceleration probability $p$.  Further parallel between
the ANS model and stochastic sandpiles are noted below.

The phase boundary shown in Fig.~\ref{tc} represents a preliminary estimate, obtained using the following criterion.  Points along the lower
critical line $p_- (\rho)$ correspond to the smallest $p$ value such that each of 200 arbitrary ICs remain active
during a time of $10^7$ steps, in a system of $L=10^{5}$ sites.  Similarly, $p_+ (\rho)$ corresponds to the largest $p$ value
such that all 200 realizations remain active.  For selected points, a precise determination was performed, as described in Sec.~\ref{critical}.
We defer a more precise mapping of the overall phase diagram to future work.

\begin{figure}[h!]
  \centering
  \includegraphics[scale=0.7]{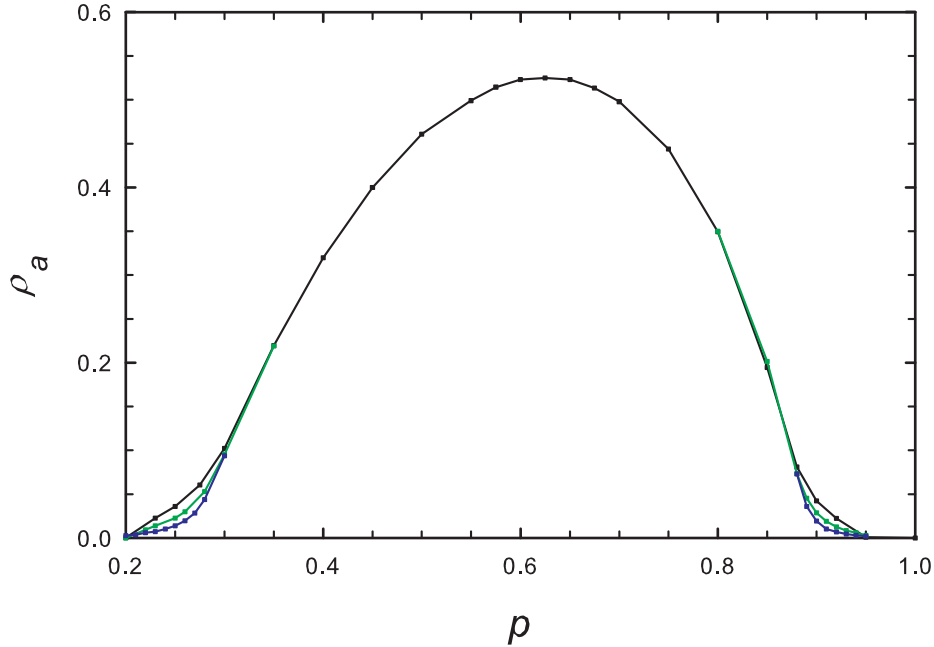}
  \caption{Steady-state activity $\rho_a$ versus $p$ for vehicle density $\rho = 1/8$. System sizes $N=1000$ (black), 2000 (green)
  and 4000 (blue). Error bars smaller than symbols.}
  \label{ravp125}
\end{figure}

The phase transitions at $p_-(\rho)$ and $p_+(\rho)$ appear to be continuous.  (Indeed, discontinuous phase
transitions between an active and an absorbing phase are not expected to occur in one-dimensional systems \cite{hinrichsen2000}.)
Figure \ref{ravp125} shows the steady-state activity (defined below) versus $p$ for density $\rho = 1/8$.  In the vicinity of the transition,
the curves become sharper with increasing system size, as expected at a continuous phase transition to an absorbing state.

\subsection{Order parameter}

Having identified a continuous absorbing-state phase transition in the ANS model, further analysis requires that we
define an appropriate order parameter or activity density.  Since the absorbing state is characterized by $v_i = v_{max}, \forall i$, one
might be inclined to define the activity density simply as $\rho_a = v_{max} - \overline{v}$.  The problem
with this definition is that not all configurations with $v_i = v_{max}, \forall i$ are absorbing: a vehicle
with $d_i = v_{max}$ may reduce its speed to $v_{max} - 1$, yielding activity in the first sense.  Since such a
reduction occurs with probability $p$, it seems reasonable to define the activity density as:

\begin{equation}
\rho_a =  v_{max} - \overline{v} + p \rho_{a,2} \equiv \rho_{a,1} + p\rho_{a,2},
\label{actdens}
\end{equation}

\noindent where $\rho_{a,2}$ denotes the fraction of vehicles with $v_i = d_i = v_{max}$.  According to this definition,
the activity density is zero if and only if the configuration is absorbing, that is, if $v_i = v_{max}$, {\it and} $d_i > v_{max}, \; \forall i$.
Studies of large systems near the critical point reveal that $\rho_{a,1} >> \rho_{a,2}$, so that the latter can be neglected
in scaling analyses.  It is nonetheless essential to treat configurations with $\rho_{a,2} > 0$ as {\it active}, even if
$\rho_{a,1} = 0$.

\subsection{Reentrance}
\label{reentrance}

In this subsection we discuss the reason for reentrance, that is, why, for $\rho_{c,<} < \rho< \rho_c$,
the system reaches the absorbing state for {\it large} $p$ as well as small $p$.  Since deceleration is associated with
generation of activity (i.e., of speeds $< v_{max}$), a reduction in activity as $p$ tends to unity seems counterintuitive.
The following intuitive argument helps to understand why this happens.
For $p \simeq 0$, vehicles rarely decelerate if they have sufficient headway to avoid reaching the position of the car in front.
This tends to increase the headway of the car behind, so that (for $\rho< \rho_c$), all headways attain values $\geq v_{max} + 1$, which
represents an absorbing configuration.  For $p=1$, a car with speed $v_i = d_i$ always decelerates, which tends to increase its own
headway.  In either case, $p=0$ or $p=1$, as reduced headway (i.e., inter-vehicle intervals with $d_i < v_{max} + 1$) is transferred down the line, vehicles may be obliged to decelerate, until
the reduced headway is transferred to an interval with headway $d_i$ large enough that no reduction in velocity is required.
[Intervals with $d_i > v_{max} + 1$, which we call {\it troughs}, always exist for $\rho < \rho_c = 1/(v_{max} + 2)$]. When all reduced headways
are annihilated at troughs, the system attains an absorbing configuration.

Call events in which a vehicle having $v_i = d_i$ decelerates {\it D events},
and those in which such a vehicle does not decelerate {\it N events}.  For $\rho < \rho_c$,
if only D events (or only N events) are allowed, the system attains an absorbing configuration via annihilation of reduced headways with troughs.
Thus some alternation between D and N events is required to maintain activity, and the active phase
corresponds to intermediate values of $p$.

\begin{figure}[h!]
  \centering
  \includegraphics[scale=0.9]{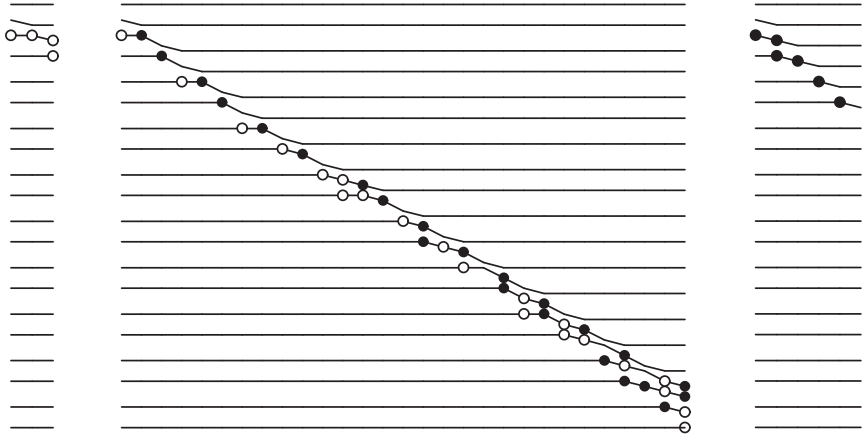}
  \vspace{-2cm}

  \caption{\label{evall} Vehicle positions relative to the first (lowest) vehicle versus time $t$ (horizontal)
  for $t \geq 2$, in a system with $N=20$, $v_{max}=2$ and vehicle density $\rho = 2/9 < \rho_c = 0.25$.  Initially,
  all vehicles have $v_i = v_{max}$.  The headways $d_i$ initially alternate between three and four,
  except for $d_{19}=0$ and $d_{20}=7$.  Filled (open) circles denote D (N) events, i.e.,
  events in which a vehicle with speed $v_i = d_i$ decelerates (does not decelerate).
  In an absorbing configuration all velocities are equal, yielding a set of horizontal lines.
  Left panel: $p=0$, system inactive for $t>4$;
  right panel: $p=1$, system inactive for $t>7$; center panel: example of a realization with $p=0.6$ in which
  activity persists until $t=56$ (evolution for $t>30$ not shown).}
\end{figure}

These observations are illustrated in Fig.~\ref{evall}, for a system of twenty vehicles with $v_{max}=2$ and density
$\rho = 2/9 < \rho_c = 0.25$.  Initially, all vehicles have $v_i = v_{max}$.  The headways $d_i$ initially alternate
between three and four (the latter are troughs),
except for $d_{19}=0$ and $d_{20}=7$.  In the left panel, for $p=0$, the system reaches an absorbing configuration after
four time steps.  Similarly, in the right panel, for $p=1$, an absorbing configuration is reached after 7 steps.  For $p=0.6$
(middle panel), the evolution is stochastic.  Most realizations reach an absorbing configuration rapidly, but some remain active
longer, as in the example shown here.  From the distribution of D and N events, it appears that activity persists
when vehicles first suffer an N event, reducing their own headway, and subsequently (one or two steps later)
suffer a D event, reducing the headway of the preceding vehicle.  Such an alternation of N and D events allows a region
with reduced headways to generate more activity before reaching a trough.

\section{Critical behavior}
\label{critical}

We turn now to characterizing the phase transition along the lines $p_-(\rho)$ and
$p_+ (\rho)$.  Since the transition is continuous, this requires that we determine the associated
critical exponents, in order to identify the universality
class of the ANS model.  The analysis turns
out to be complicated by strong finite-size effects: different from simple systems exhibiting an
absorbing-state phase transition, such as the contact process, for which studies of systems with $L \leq 1000$
yield good estimates for critical exponents \cite{marro}, here we require systems of up to $10^5$ sites
to obtain reliable results.
We are nevertheless able to report
precise results at several points along the phase boundary.

We use quasistationary (QS) simulations to probe the behavior at long times conditioned
on survival of activity \cite{qssim}.  Since the deceleration probability $p$ is continuous while the density $\rho$ can only be
varied in discrete steps, we keep the latter fixed and vary the former in each series
of studies.
As in other studies of QS behavior absorbing-state phase transitions, we focus on the
finite-size scaling (FSS) of the activity density, $\rho_a$, the lifetime,
$\tau$, and the moment ratio $m = \langle \rho_a^2 \rangle/\rho_a^2$, as functions of
system size, $N$ \cite{marro,qssim}.  At a critical point, these variables are expected to exhibit scale-free (power-law)
dependence on $N$, that is, $\rho_a \sim N^{-\beta/\nu_\perp}$
and $\tau \sim N^z$, where $\beta$ is the order-parameter exponent and $\nu_\perp$ the exponent that governs the
divergence of the correlation length as one approaches the critical point.
In the active phase, $\rho_a$ approaches a nonzero constant value, while $\tau$ grows exponentially as $N \to \infty$.
In the absorbing phase, $\rho_a \sim 1/N$ while $\tau$ grows more slowly than a power law as $N \to \infty$.
At the critical point,
the moment ratio is expected to converge to a nontrivial limiting value, $m = m_\infty + {\cal O} (N^{-\lambda})$,
with $\lambda > 0$.  In the active (inactive) phase, $m$ curves sharply downward (upward) when plotted
versus $1/N$.  These are the criteria we employ to determine the critical point, $p_c(\rho)$.
The distance from the critical point can be estimated from the curvature of log-log plots of $\rho_a$
and $\tau$ versus $N$.

As noted in Sec. III.B, the order parameter is the sum of two contributions: $\rho_a = \rho_{a,1} + p\rho_{a,2}$.
In simulations, we therefore determine $\rho_{a,1}$ and $\rho_{a,2}$ separately.  In the vicinity of the critical point we find
$\rho_{a,1} \sim N^{-0.5}$ and $\rho_{a,2} \sim N^{-0.9}$, showing that the fraction $\rho_{a,2}$ of vehicles with $v_i = d_i = v_{max}$
decays more rapidly than $\rho_{a,1} = v_{max} - \overline{v}$, so that it makes a negligible contribution to
the activity density for large $N$.  We therefore adopt $\rho_{a,1}$ as the order parameter
for purposes of scaling analysis.
Configurations $\rho_{a,1}=0$ and $\rho_{a,2} > 0$ are nevertheless considered to be active; only configurations with
$v_i = v_{max}$ and $d_i > v_{max}$, $\forall i$, are treated as absorbing.

We study rings of 1000, 2000, 5000, 10$\,$000, 20$\,$000, 50$\,$000 and 100$\,$000 sites,
calculating averages over a set of 20 to 160 realizations.
Even for the largest systems studied, the activity density reaches
a stationary value within $10^6$ time steps.  We perform averages over the subsequent
$10^8$ steps.  As detailed in \cite{qssim},
the QS simulation method probes the quasistationary probability
distribution by restarting the evolution in a randomly chosen active configuration
whenever the absorbing state is reached.  A list of $N_c$ such configurations, sampled from the
evolution, is maintained; this list is renewed by exchanging one of the saved configurations with the
current one at rate $p_{r}$.  Here we use $N_c = 1000$, and $p_r = 20/N$.
During the
relaxation phase, we use a value of $p_r$ that is ten times greater, to eliminate the vestiges of the
initial configuration from the list.
The lifetime $\tau$ is taken as the mean time
between attempts to visit an absorbing configuration, in the QS regime.

Initial configurations are prepared by placing vehicles as uniformly as possible (for example, for
density $\rho = 1/8$, we set $d_i = 7$, $\forall i$), and then exchanging distances randomly.
In such an exchange a site $j$ is chosen at random and the changes $d_j \to d_j -1$ and
$d_{j+1} \to d_{j+1} + 1$ are performed, respecting the periodic boundary condition, $d_{N+1} \equiv d_{1}$.
The random exchange is repeated $N_e$ times (in practice
we use $N_e = 2N$), avoiding, naturally, negative values of $d_{j}$. Since headways $d_j < v_m$ are generated in this process,
at the first iteration of the dynamics, velocities $v_j <  v_{max}$ arise, leading to a relatively large, statistically
uniform initial activity density.

\begin{figure}[h!]
  \centering
  \includegraphics[scale=0.7]{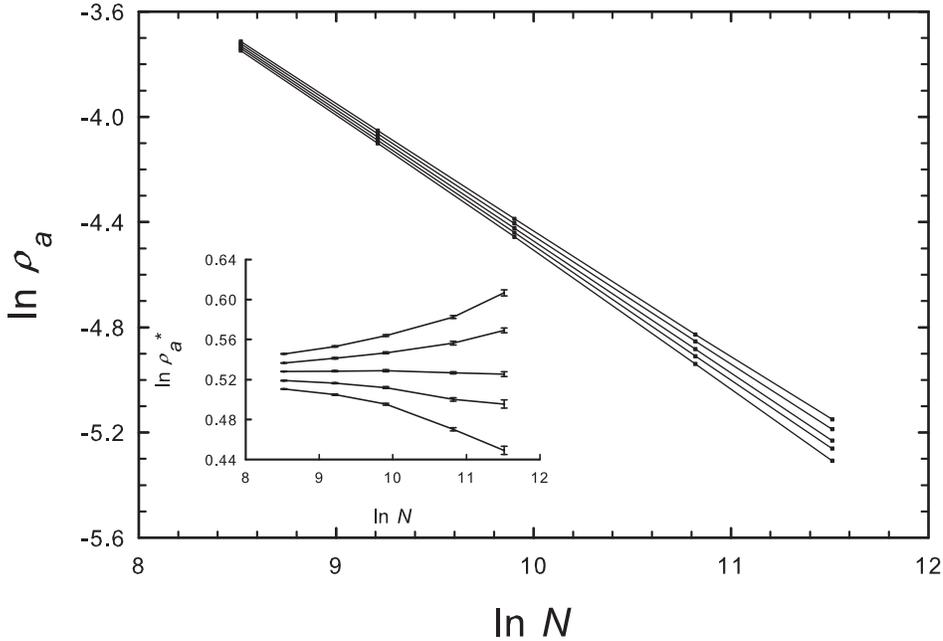}
  \caption{\label{rhoavN} Activity density versus number of vehicles for density 1/8 and (lower to upper)
  $p = 0.2679$, 0.2681, 0.2683, 0.2685 and 0.2687.  Error bars are smaller than symbols. Inset:
  scaled activity density $\rho_a^* = N^{0.5} \rho_a$ versus number of vehicles.}
\end{figure}

\begin{figure}[h!]
  \centering
  \includegraphics[scale=0.7]{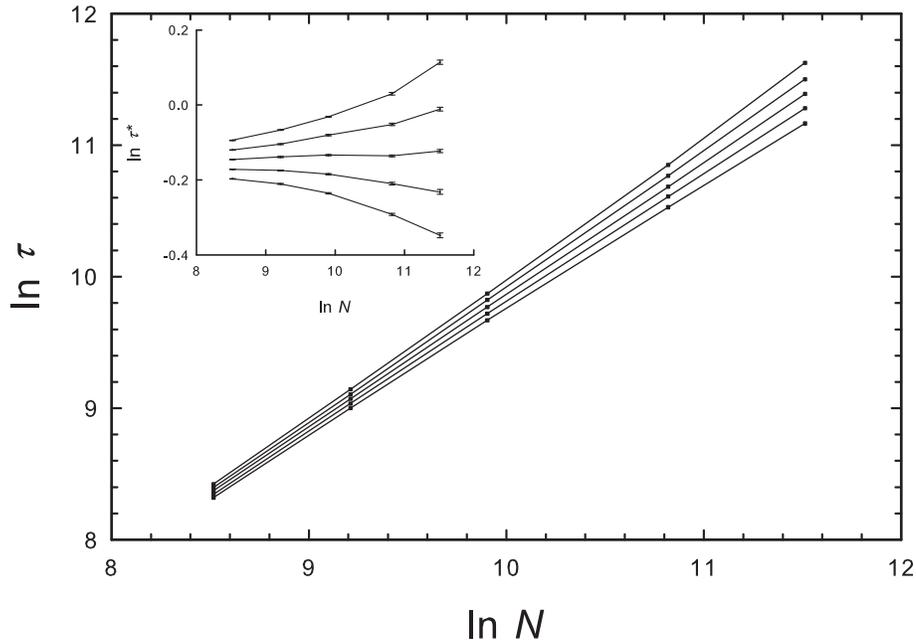}
  \caption{\label{tauvN} Lifetime versus number of vehicles for density 1/8 and (lower to upper)
  $p = 0.2679$, 0.2681, 0.2683, 0.2685 and 0.2687.  Error bars are smaller than symbols. Inset:
  scaled lifetime $\tau^* = N^{-1.0} \tau$ versus number of vehicles.}
\end{figure}

\begin{figure}[h!]
  \centering
  \includegraphics[scale=0.7]{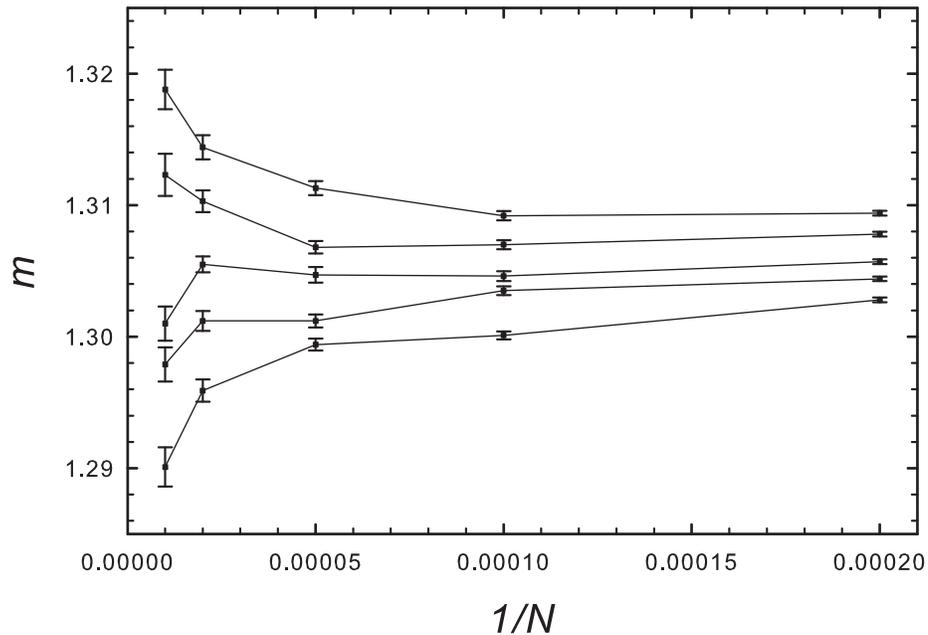}
  \caption{\label{mvrN} Moment ratio $m$ versus reciprocal system size for density 1/8 and (upper to lower)
  $p = 0.2679$, 0.2681, 0.2683, 0.2685 and 0.2687.}
\end{figure}

We performed detailed studies for densities $\rho = 1/8$, on both the upper and lower critical lines, and for density
17/144 $= 0.1180\overline{5}$, on the lower line.
Figures \ref{rhoavN}, \ref{tauvN} and \ref{mvrN} show, respectively, the dependence of the order parameter, lifetime and moment ratio $m$
on system size for density 1/8 and $p$ values in the vicinity of the lower critical line.
In the insets of Figs. \ref{rhoavN} and \ref{tauvN} the values of $\rho_a$ and $\tau$ are divided by the overall
trend to yield $\rho_a^* \equiv N^{0.5} \rho_a$ and $\tau^* = \tau/N$. These plots make evident subtle
curvatures hidden in the main graphs, leading to the conclusion that $p_c (\rho = 1/8)$ is very near 0.2683.

\begin{figure}[h!]
  \centering
  \includegraphics[scale=0.7]{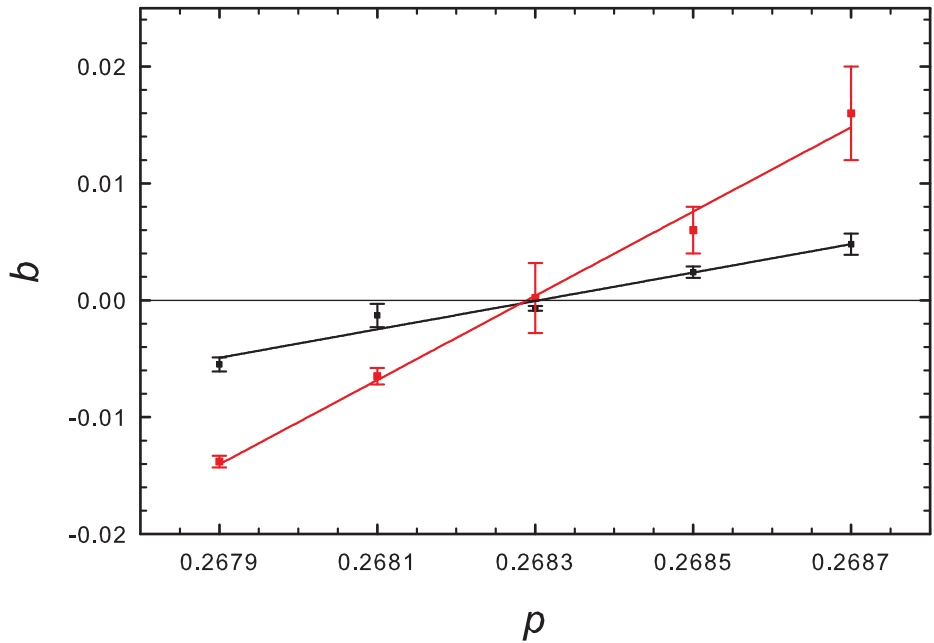}
  \caption{\label{bvp} (Color online) Curvature of $\ln \rho_a$ (black points) and $\ln \tau$ (red points)
  as functions of $\ln N$, as measured by the coefficient $b$ of the quadratic term in least-squares quadratic
  fits to the data in Figs.~\ref{rhoavN} and \ref{tauvN}.  Straight lines are least-squares linear fits to
  $b$ versus deceleration probability $p$, for vehicle density $\rho = 1/8$.  Intercepts with the line $b=0$
  furnish estimates of $p_c$.}
\end{figure}

A more systematic analysis involves the curvatures of these quantities: we fit quadratic polynomials,

\begin{equation}
\ln \rho_a = \mbox{const.} + a \ln N + b (\ln N)^2,
\end{equation}

\noindent and similarly for $\ln \tau$,
to the data for the four largest system sizes.
The coefficient of the quadratic term, which should be zero at the critical point, is plotted versus $p$ in Fig.~\ref{bvp}.
Linear interpolation to $b=0$ yields
the estimates $p_c = 0.26830(3)$ (data for activity density) and $p_c = 0.26829(2)$ (data for lifetime);
we adopt $p_c = 0.26829(3)$ as our final estimate.  (Figures in parentheses denote statistical uncertainties.)
The data for $m$, although more scattered, are consistent with this estimate: from Fig.~\ref{mvrN} it is evident that
$p_c$ lies between 0.2681 and 0.2683.

To estimate the critical exponents $\beta/\nu_\perp$ and $z$ we perform linear fits to the data for $\ln \rho_a$
and $\ln \tau$ versus $\ln N$ (again restricted to the four largest $N$ values), and consider the slopes
as functions of $p$.  Interpolation to $p_c$ yields the estimates: $\beta/\nu_\perp = 0.500(3)$ and $z = 1.006(8)$.
A similar analysis yields $m_c = 1.306(6)$.
The principal source of uncertainty
in these estimates is the uncertainty in $p_c$.

\begin{figure}[h!]
  \centering
  \includegraphics[scale=0.7]{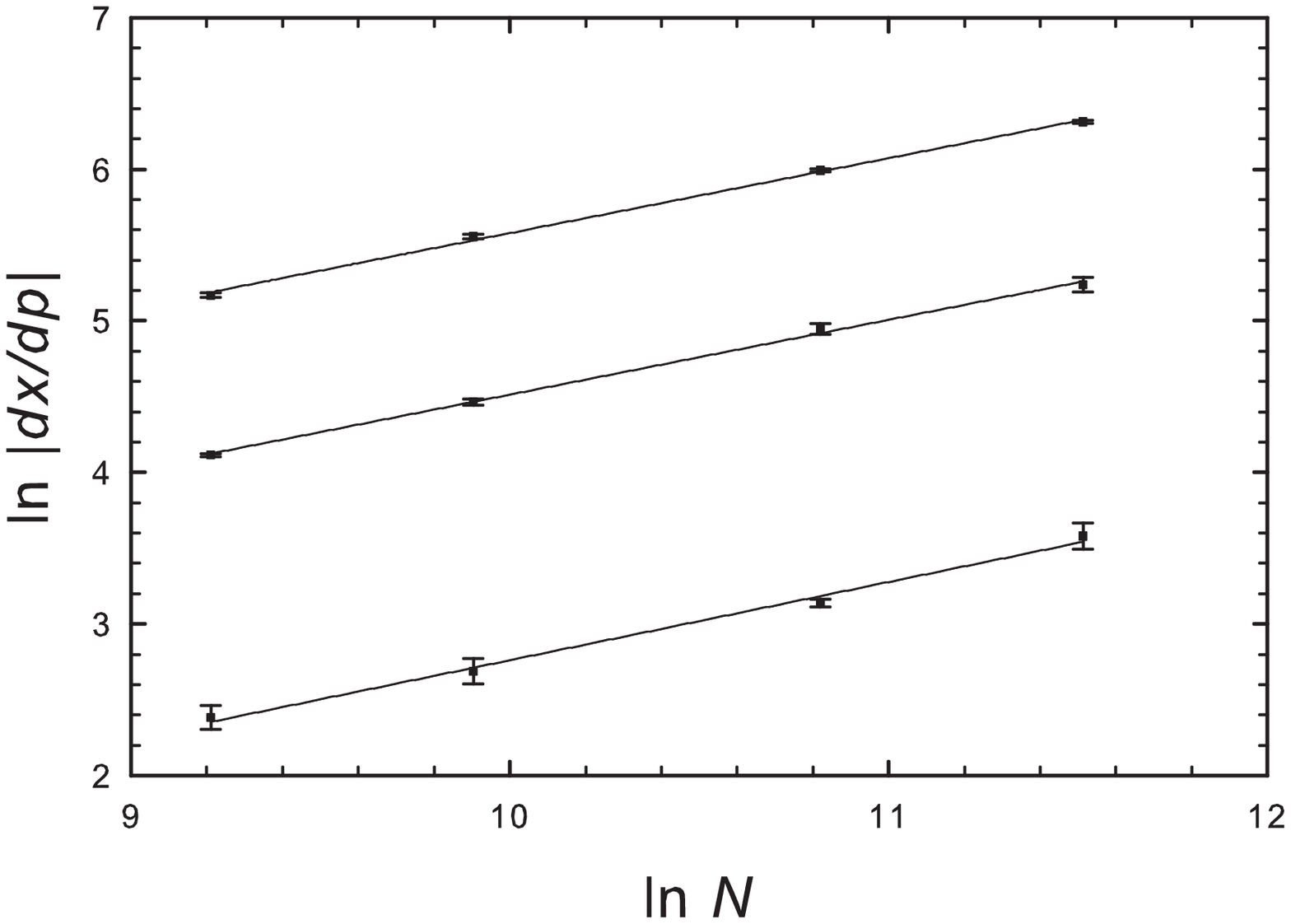}
  \caption{\label{derivs} Derivatives of (lower to upper) $m$, $\ln \rho_a$ and $\ln \tau$ with respect to $p$
  in the vicinity of $p_c$, versus $N$ for vehicle density $\rho = 1/8$.  Lines are least-squares linear
  fits to the data.}
\end{figure}

Using the data for $\rho_a$, $\tau$ and $m$ we also estimate the critical exponent $\nu_\perp$.
Finite-size scaling implies that the derivatives
$|dm/dp|$, $d \ln \tau/dp$ and $d \ln \rho_a/dp$, evaluated at the critical point,
all grow $ \propto L^{1/\nu_\perp}$ .
We estimate the derivatives via least-squares linear fits to the data on an interval
that includes $p_c$. (The intervals are small enough that the graphs show no significant
curvature.)  Power-law dependence of the derivatives on system size is verified in
Fig.~\ref{derivs}.  Linear fits to the data for the four largest sizes, for $\ln \rho_p$,
$\ln \tau$, and $m$ yield
$1/\nu_\perp = 0.494(15)$, 0.495(15), and 0.516(29), respectively, leading to the estimate
$\nu_\perp = 2.00(5)$.
Repeating the above analysis for simulations at vehicle density $\rho = 17/144$, we find $p_-(17/144) = 0.4096(1)$,
$\beta/\nu_\perp = 0.503(6)$, $z = 1.011(15)$, $m=1.302(2)$, and $\nu_\perp = 2.02(3)$.

Thus, for the two points studied on the lower critical line, the results are consistent
with a simple set of exponent values, namely, $z=1$, $\nu_\perp = 2$, and $\beta=1$.
The same set of critical exponents appears in a system of activated random walkers (ARW) on a ring,
when the walkers hop in one direction only \cite{arw}.  The critical moment ratio for ARW is $m_c = 1.298(4)$, quite near present estimates.
We suggest that these values characterize a universality class of absorbing-state phase transitions
in systems with a conserved density (of walkers in ARW, and of vehicles in the present instance), and
anisotropic movement.  The ARW with {\it symmetric} hopping is known to belong to the universality class
of conserved directed percolation \cite{slwkr}, which also includes conserved stochastic sandpiles \cite{granada,pruessner}.

A study on the upper critical line for vehicle density $\rho = 1/8$ yields results that are similar but
slightly different.  Repeating the procedure described above, we find $p_+ (1/8) = 0.89590(5)$,
$\beta/\nu_\perp = 0.487(8)$, $z = 1.021(15)$, $\nu_\perp = 1.98(6)$, and $m_c = 1.315(5)$.  The exponent values are sufficiently near
those obtained on the lower critical line that one might attribute the differences to
finite-size effects.  We defer to future work more detailed analyses, to determine whether scaling properties
along the upper and lower critical lines differ in any respect.

\section{Summary}

We consider a version of the Nagel-Schreckenberg model in which probabilistic deceleration is possible only
for vehicles whose velocity is equal to the headway, $v_i = d_i$.
In the resulting ANS model, a free-flow configuration, $v_i = v_{max}$ and $d_i > v_{max}, \; \forall i$,
is {\it absorbing} for any value of the deceleration probability $p$.
The phase transition in the original NS model at deceleration probability $p=0$ is identified with
the absorbing-state transition in the ANS model: the two models are identical for $p=0$.
In the original model, a nonzero deceleration probability corresponds to a spontaneous source of activity
which eliminates the absorbing state, and along with it, the phase transition.

The ANS model, by contrast, exhibits a line of absorbing-state phase transitions in the $\rho$-$p$ plane; the phase
diagram is reentrant. We present preliminary estimates for the phase boundary and several critical exponents.
The latter appear to be associated with a universality class of absorbing-state phase transitions
in systems with a conserved density and asymmetric hopping, such as activated random walkers (ARWs) with particle
transfer only in one direction \cite{arw}. In this context it is worth noting that in traffic models, as well as in
sandpiles and ARW, activity is associated with a local excess of density: in sandpiles, activity requires sites with
an above-threshold number of particles; in ARW, it requires an active particle jumping to a site
occupied by an inactive one; and in the ANS model, it requires headways $d$ smaller than $v_{max}+1$.  One may hope that
the connection with stochastic sandpiles will lead to a better understanding of traffic models, and perhaps of
observed traffic patterns.

\begin{acknowledgments}
This work was supported by CNPq and CAPES, Brazil.
\end{acknowledgments}

\end{document}